\documentclass[lettersize,journal]{IEEEtran}

\usepackage{amsmath,amssymb,amsfonts} % math
\usepackage{booktabs, makecell, multirow} % cells in tables
\usepackage{bigdelim} % curly braces in table
\usepackage{acronym} % used in the text are spelled out in full at least once. 
\usepackage{cite}
\usepackage{url}
\usepackage{tabularray}
\UseTblrLibrary{booktabs}
\usepackage{bigdelim}
\usepackage{blindtext}
\usepackage{graphicx}
\usepackage{adjustbox}
\graphicspath{{./figures}}
\usepackage{siunitx}
\usepackage{float}
\usepackage{hyperref}
\usepackage{bm} 
\usepackage{enumitem}
\usepackage{bbding}
\usepackage[dvipsnames]{xcolor} 
\acrodef{STFT}{short-time Fourier transform}
\acrodef{ISTFT}{inverse short-time Fourier transform}
\acrodef{BSS}{blind source separation}
\acrodef{DOA}{direction of arrival}
\acrodef{DC}{deep clustering}
\acrodef{DPRNN}{dual-path recurrent neural network}
\acrodef{TF}{time-frequency}
\acrodef{TCN}{temporal convolutional network}
\acrodef{ATF}{acoustic transfer function}
\acrodef{SI-SDR}{scale-invariant signal-to-distortion ratio}
\acrodef{MSE}{mean square error}
\acrodef{DFT}{discrete Fourier transform }
\acrodef{SIR}{signal-to-interference ratio}
\acrodef{OVA}{overlap-and-add}
\acrodef{SDR}{signal-to-distortion ratio}
\acrodef{BLSTM}{Bidirectional Long Short-Term Memory}
\acrodef{SOTA}{state-of-the-art}
\acrodef{RI}{Real-Imaginary}
\acrodef{RIR}{room impulse response}
\acrodef{SNR}{signal-to-noise ratio}
\acrodef{RNN}{Recurrent Neural Networks}
\acrodef{FLOP}{floating point operation}
\acrodef{WPE}{weighted prediction error}
\acrodef{CASA}{computational simultaneous grouping scene analysis}
\acrodef{FC}{fully connected}
\acrodef{E2E}{end-to-end}
\acrodef{SDR}{signal to distortion ration}
\acrodef{SIR}{signal-to-interference ratio}
\acrodef{STOI}{short-time objective intelligibility}
\acrodef{PESQ}{perceptual evaluation of speech quality}
\acrodef{CNN}{convolutional neural network}  % for acronym

\usepackage{balance}
\begin{document}
\title{Typing to Listen at the Cocktail Party: Text-Guided Target Speaker Extraction}

% \author{Xiang Hao$^{1,2}$, Jibin Wu$^{1}$\thanks{Corresponding author}, Jianwei Yu$^{2}$, Chenglin Xu$^{1}$ \& Kay Chen Tan$^{1}$ \\
% ${^1}$Department of Computing, The Hong Kong Polytechnic University\\
% ${^2}$Tencent AI Lab\\
% \texttt{haoxiangsnr@gmail.com, jibin.wu@polyu.edu.hk} \\
% % \texttt{\href{https://github.com/haoxiangsnr/llm-tse}{https://github.com/haoxiangsnr/llm-tse}}
% }

\author{Xiang~Hao,
        Jibin~Wu,
        Jianwei~Yu,
        Chenglin~Xu,
        Kay~Chen~Tan~\IEEEmembership{Fellow,~IEEE}% <-this % stops a space
\thanks{Xiang Hao, Jibin Wu, Chenglin Xu and Kay Chen Tan are with the Department of Computing, The Hong Kong Polytechnic University, Hong Kong SAR, China.}% 
\thanks{Jianwei Yu is with Tencent AI Lab.}
\thanks{Jibin Wu (jibin.wu@polyu.edu.hk) is the corresponding author.}
}


\markboth{Journal of \LaTeX\ Class Files,~Vol.~14, No.~8, August~2015}%
{Shell \MakeLowercase{\textit{et al.}}: Bare Demo of IEEEtran.cls for IEEE Journals}


\maketitle

\begin{abstract}
Humans can easily isolate a single speaker from a complex acoustic environment, a capability referred to as the ``Cocktail Party Effect." However, replicating this ability has been a significant challenge in the field of target speaker extraction (TSE). Traditional TSE approaches predominantly rely on voiceprints, which raise privacy concerns and face issues related to the quality and availability of enrollment samples, as well as intra-speaker variability. To address these issues, this work introduces a novel text-guided TSE paradigm named LLM-TSE. In this paradigm, a state-of-the-art large language model, LLaMA 2, processes typed text input from users to extract semantic cues. We demonstrate that textual descriptions alone can effectively serve as cues for extraction, thus addressing privacy concerns and reducing dependency on voiceprints. Furthermore, our approach offers flexibility by allowing the user to specify the extraction or suppression of a speaker and enhances robustness against intra-speaker variability by incorporating context-dependent textual information. Experimental results show competitive performance with text-based cues alone and demonstrate the effectiveness of using text as a task selector. Additionally, they achieve a new state-of-the-art when combining text-based cues with pre-registered cues. This work represents the first integration of LLMs with TSE, potentially establishing a new benchmark in solving the cocktail party problem and expanding the scope of TSE applications by providing a versatile, privacy-conscious solution. Demos are provided at \href{https://github.com/haoxiangsnr/llm-tse}{https://github.com/haoxiangsnr/llm-tse}\footnote{The source code and datasets will be made publicly available after review.}
\end{abstract}

% These cues can independently direct the TSE process, serve as task selectors, or augment pre-registered cues. 
\begin{IEEEkeywords}
target speaker extraction, speaker separation, large language models, speech signal processing, audio-text multimodal modeling
\end{IEEEkeywords}

% \fi
\IEEEpeerreviewmaketitle

\section{Introduction}
\label{sec:intro}

\IEEEPARstart{H}{umans} have an innate ability to focus on a specific single auditory source while filtering out other undesired auditory sources or background noise, which is referred to as the ``Cocktail Party Effect"~\cite{cherry1953some}.
This human skill, though seemingly effortless, acutally conceals the complexity that has long challenged scientists and engineers in their quest to replicate it artificially~\cite{haykin_cocktail_2005,mesgarani_selective_2012,bizley_what_2013,gannot_consolidated_2017}.
In the domain of computational auditory scene analysis, target speaker extraction (TSE) ~\cite{zmolikova_speakerbeam_2019,xu_spex_2020,chen_continuous_2020,raj_integration_2021,yoshioka_recognizing_2018,zmolikova_neural_2023} has been a focal point of research, which isolates a specific speaker's voice from a mixture of sounds.
Recent previous TSE approaches mainly employ voiceprints to discern and isolate the speaker's voice from a mixture signal, which are extracted from pre-recorded enrollment utterances with computational models like Convolutional Neural Networks (CNNs)~\cite{xu_spex_2020,ohishi_conceptbeam_2022,ge2022spex}, Recurrent Neural Networks (RNNs)~\cite{zmolikova_speakerbeam_2019,luo_speaker-independent_2018}, and Transformers~\cite{avsepformer23,transformerrealtime23}. 
Despite their remarkable effectiveness, these approaches face significant challenges.
\textbf{1) Privacy concerns.} Privacy concerns are at the forefront of public discourse, especially when it involves the use of a speaker's voice~\cite{Alegre2009}. Voiceprint-based extraction systems necessitate the collection of a sample voice for enrollment purposes. This requirement raises privacy issues that can greatly limit the adoption and practicality of TSE systems.
\textbf{2) Availability of high-quality cues.} Even with user consent, the availability of high-quality, lengthy pre-recorded enrollment speech is not guaranteed. Challenges including inconsistent recording channels, pervasive background noise, and inadequate sample duration can significantly degrade the performance of TSE systems~\cite{zmolikova_speakerbeam_2019,xu_spex_2020,jianwei2023tse,zmolikova_neural_2023}.
\textbf{3) Intra-speaker variability.} Even with access to high-quality enrollment speech of sufficient length, the speech signal of the same speaker might have highly different characteristics in different conditions due to such factors as acoustic environment (e.g., different room geometry structures or microphone frequency responses) or emotional state (e.g., happy, sad, or angry). It is very challenging to make TSE systems robust enough to such intra-speaker variability~\cite{zmolikova_neural_2023}.

Given these hurdles, we turn back to the innate human ability to identify and describe the target speaker succinctly and effectively, such as requesting to ``Extract the speaker who is saying `Paris 2024 Summer Olympics' from the audio," or ``Extract the loudest speaker from the mixture."
This method of describing the target speaker through natural language is not only straightforward and cost-effective but also privacy-conscious and does not require professional recording equipment, while still offering discriminability. 
To perform such human-like target speaker extraction, an essential prerequisite is to make machines well understand the auditory object perception differences described by humans in natural language. 
To date, this has been feasible with the significant advancements made by large language models (LLMs)~\cite{li_videochat_2023,zhang_speechgpt_2023,huang_audiogpt_2023}, which have demonstrated amazing capabilities of natural language understanding. 

\begin{figure*}[ht]
    \centering
    \includegraphics[width=1\textwidth]{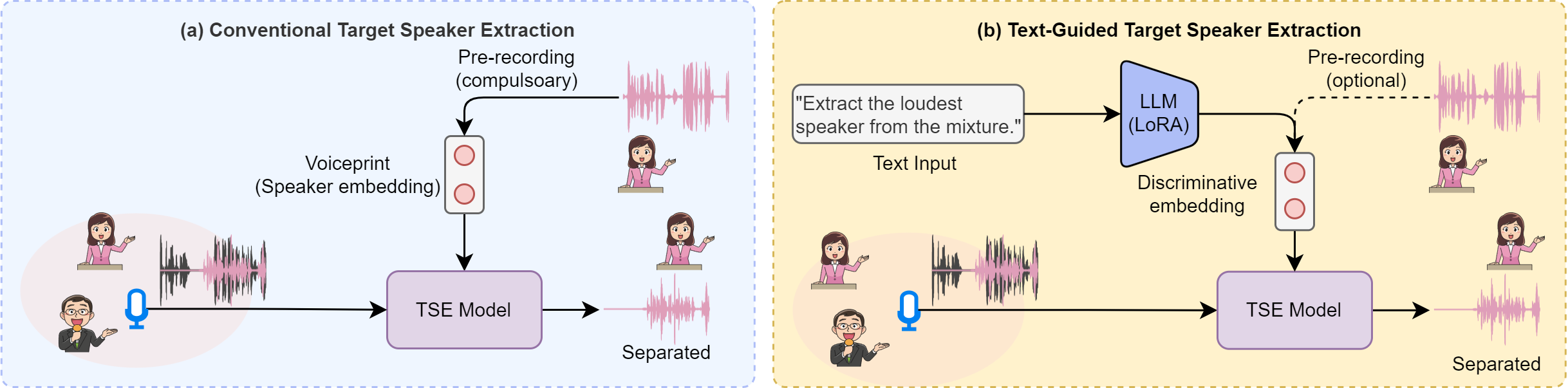}
    \caption{Comparison between conventional TSE system and our proposed Text-Guided TSE system. The former relies on the pre-registered voiceprint of the target speaker as an extraction cue, while our system offers flexibility to incorporate text-based cues to facilitate target speaker extraction.}
    \label{fig:diff}
\end{figure*}

Hence, we develop an innovative text-guided target speaker extraction paradigm, named LLM-TSE, as depicted on Figure~\ref{fig:diff} (b). 
LLM-TSE employs a text encoder based on a state-of-the-art LLM to interpret user-provided natural language descriptions, thereby isolating the speech signal of a target speaker from a mixture of several speakers.
It provides a novel solution that can function independently or complement traditional techniques for the TSE tasks especially when conventional cues like voiceprints are unavailable or impossbile to access. 
Specifically, the proposed LLM-TSE consists of three main modules: a text cue encoder, an audio cue encoder, and a speech extraction module. 
The text cue encoder leverages the strong understanding capabilities~\cite{Wei2022EmergentAO,KASNECI2023102274} of the state-of-the-art LLM model LLaMA 2~\cite{touvron_llama_2023} to interpret natural language text descriptions and extract semantic cues that inform the target speaker extraction process.
These descriptions cover various aspects of human auditory perception, including speaker characteristics, language, conversation content, room characteristics, and more.
An optional audio cue encoder is employed to utilize the enrollment speech of the target speaker when available.
These two cues can work independently or even simultaneously.
For example, given a pre-recorded enrollment voice, users can tell the model to ``eliminate the target speaker's voice" rather than extracting it, or further inform the model of the current state of the target speaker using the text like ``the target speaker is the near-field speaker in the audio".
Finally, the speech extraction module estimates the target speech from the mixture utilizing the target speaker embedding derived from the cues provided.

The proposed text-based approach offers several advantages: \textbf{1) Privacy-friendliness.} Unlike voiceprints, text does not necessarily carry personally identifiable information, making it a more acceptable option in terms of privacy protection. 
\textbf{2) Cost-efficiency.} Text is undoubtedly less expensive compared to other forms of cues such as target voices, angles, images, and videos. 
\textbf{3) Flexibility.} The use of text allows for selectively retaining or removing the source of interest based on the semantic concepts expressed in the text. Using text as a control mechanism, the system becomes a unified and flexible approach that avoids the need for training multiple systems. 
\textbf{4) Contextual robustness.} Textual input enables us to inform the model of the speaker's current state (including acoustic environment and speaker state) to help tackle intra-speaker variability. Additional cues that align with human perception of speech mixtures are incorporated to lift the effectiveness of TSE in practical scenarios.

We conduct extensive experiments on the mixture overlapped speech dataset, and it has been well demonstrated that our proposed method achieves performance comparable to that of the audio-only systems when relying solely on text input. 
When audio cues are available, text input can effectively serve as a task selector, accurately determining the type of task at hand. 
Furthermore, when text is utilized to provide additional information about the current state of a speaker with a pre-recorded enrollment speech, the model's performance significantly exceeds that of the audio-only extraction systems.

To the best of our knowledge, this is the first study to utilize natural language descriptions for target speaker extraction. The contributions of this work are threefold:
\begin{itemize}[leftmargin=0.3cm]
    \item This work pioneers the use of natural language descriptions as standalone cues for target speaker extraction, showcasing their efficacy and addressing privacy concerns associated with voiceprint-based approaches.
    \item This work introduces a flexible control mechanism via natural language input, simplifying the speaker extraction process and enhancing the system's adaptability across various scenarios.
    \item This work combines context-dependent information from text with traditional cues, offering a robust solution to intra-speaker variability and improving the practicality of speaker extraction systems.
\end{itemize}

The remainder of this work is structured as follows: A discussion of works related to our research is presented in Section~\ref{sec:others}. Section~\ref{sec:new_app} provides an overview of novel application scenarios enabled by the proposed LLM-TSE model. Section~\ref{sec:model} delineates the intricate architecture of the LLM-TSE model. The experimental setup and corresponding results are detailed in Section~\ref{sec:exp_setup} and Section~\ref{sec:exp_result}, respectively. Finally, Section~\ref{sec:con} concludes the paper by summarizing our findings and outlining avenues for future investigation.

\section{Related Works}
\label{sec:others}

\subsection{Speech Separation and Target Speaker Extraction}
To solve the Cocktail Party problem, early research efforts mainly adopt computational auditory scene analysis (CASA)~\cite{lyon1983computational,meddis1991virtual,seltzer2003harmonic,wang2006computational}, non-negative matrix factorization (NMF) \cite{cichocki2006new,virtanen2007monaural,parry2007incorporating}, and factorial Hidden Markov Models and Gaussian Mixture Models (HMM-GMM) \cite{virtanen2006speech,stark2011source}. 
These methods are often limited by the representation power of their models, resulting in poor performance in complex acoustic environments. 
In recent decades, the advent of deep learning has significantly advanced the progress in this field. 
Existing DNN-based techniques can be broadly classified into two categories: blind source separation (BSS)~\cite{pal_blind_2013,hershey_deep_2016,yu_permutation_2017,luo_conv-tasnet_2019} and target speaker extraction (TSE)~\cite{luo_speaker-independent_2018,zmolikova_speakerbeam_2019,xu_spex_2020,ge_spex_2020,pan2022selective,zmolikova_neural_2023}. 
BSS techniques usually adopt DNNs to estimate an auditory mask for each speaker, which is then leveraged to separate each speaker's voice into an individual stream from the mixture speech captured by a microphone. 
A difficulty in this process stems from the global permutation ambiguity~\cite{hershey_deep_2016}, which hampers the assignment of the output of a multi-source separation system to the correct source accurately. 
To address it, deep clustering (DC) techniques~\cite{hershey_deep_2016,isik_single-channel_2016,wang_alternative_2018} are proposed to group the spectro-temporal features belonging to the same speaker through a clustering scheme. 
Permutation invariant training (PIT)~\cite{yu_permutation_2017,kolbaek_multitalker_2017} is invented to solve this problem by finding the minimal loss over all the permutations between the extracted streams and the reference speeches. 
Typically, these methods require prior knowledge or estimation of the number of speakers in the mixture. However, in real-world scenarios, the number of speakers is hard to predict in advance.

Target speaker extraction (TSE) provides an alternative solution to address the challenges of the unknown number of speakers and global permutation ambiguity.
This approach involves providing a cue that is related to the desired speaker, such as a pre-recorded speech describing the voice characteristics~\cite{zmolikova_speakerbeam_2019,xu_spex_2020}, a spatial cue indicating the speaker's direction~\cite{ge2022spex}, or synchronous lip movement~\cite{pan2022selective}. By using these specified cues, only the target speaker's voice is extracted, thereby avoiding the issue of the unknown number of speakers and global permutation ambiguity. However, efforts on developing such systems are confronted by a number of challenges as mentioned in Section~\ref{sec:intro}.

\subsection{Audio-Language Multimodal Modeling}
Audio-language multimodal modeling is currently a significant research area with many application scenarios~\cite{huang_audiogpt_2023,zhang_speechgpt_2023,gong_listen_2023}. The primary focus has revolved around audio events, with most tasks and datasets originating from automatic audio caption~\cite{drossos_automated_2017,wu_audio_2019,mei_automated_2022}, which aims to assign meaningful textual descriptions to audio content. Leveraging these datasets, related studies have been conducted on synthesizing audio based on text descriptions, which find applications in diverse scenarios such as film production, game design, and more.
Among them, the Contrastive Language-Audio Pretraining (CLAP)~\cite{elizalde_clap_2022} model is a large-scale pre-training model that employs a contrastive learning approach similar to the Contrastive Language-Image Pretraining (CLIP)~\cite{radford_learning_2021} model for aligning text and audio modalities. This model has pushed the boundaries in tasks involving synthesizing audio based on text descriptions~\cite{huang_make--audio_2023,kreuk_audiogen_2023,liu_audioldm_2023,liu_audioldm_2023-1}. 
Furthermore, the works \cite{wang_speechx_2023,zhang_speechgpt_2023,le_voicebox_2023} expand the input modality to encompass audio and text instead of text only for audio generation. However, note that the underlying logic is based on generative models that take audio and specific control inputs to handle various speech transformation tasks. These works are more like controlled speech/audio/music synthesis, not requiring the length of input and output to be strictly aligned. This is entirely different from the field of our study.

\subsection{Audio-Language-Vison Multimodal Target Source Separation} 
Among all these audio-language multimodal models, the most relevant to our research involve separating or detecting audio events based on text description~\cite{kilgour_text-driven_2022,liu_separate_2022,liu_separate_2023, li_target_2023}. These studies employ models like BERT~\cite{devlin_bert_2019} (mini) or CLAP~\cite{elizalde_clap_2022} to comprehend descriptions of sound events, subsequently separating the sound sources consistent with the target description. 
However, they are not specifically designed for speech signals. In contrast to audio event classes, speech signals are considerably similar when observed from spectrograms, lacking clear acoustic spectral patterns to follow. Instead, they rely more on perceptual differences in auditory objects and semantic information. In addition to sound events, these models also focus on separating musical instruments~\cite{chen_musicldm_2023,huang_noise2music_2023,chen_musicldm_2023}.
While these previous works have made big strides, the specific challenges and nuances of speech signal separation are out of their scope.
Labels, particularly those implemented via one-hot vectors~\cite{potdar2017comparative}, can be seen as a distinctive type of human language.
In the realm of label-based audio/music/speech extraction systems~\cite{manilow_hierarchical_2020,delcroix_few-shot_2021,tzinis_heterogeneous_2022,delcroix_soundbeam_2023, li_target_2023, ochiai_listen_2020}, the works of \cite{manilow_hierarchical_2020} and \cite{tzinis_heterogeneous_2022} are most closely aligned with ours. These systems, like ours, endeavor to integrate human subjective intentions into the separation process through attribute labels. Yet, they solely rely on one-hot vectors, resulting in a lack of flexibility within human-computer dialogue systems. In addition, they cannot understand the vast array of human language inputs and struggle significantly when dealing with open-ended queries. 
By contrast, we employ LLMs to understand human descriptions of auditory object differences, which offers increased flexibility in cue extraction.
Furthermore, we investigate control capabilities of human descriptions and explore combining cues of the text-and-audio multimodal input.
Another related method utilizes semantic cues, i.e. images~\cite{ohishi_conceptbeam_2022}, to extract speakers' speech discussing a particular concept. However, Finding the right images as cues for extraction is very hard and expensive in practice.

\begin{figure*}[!t]
    \centering
    \includegraphics[width=0.95\textwidth]{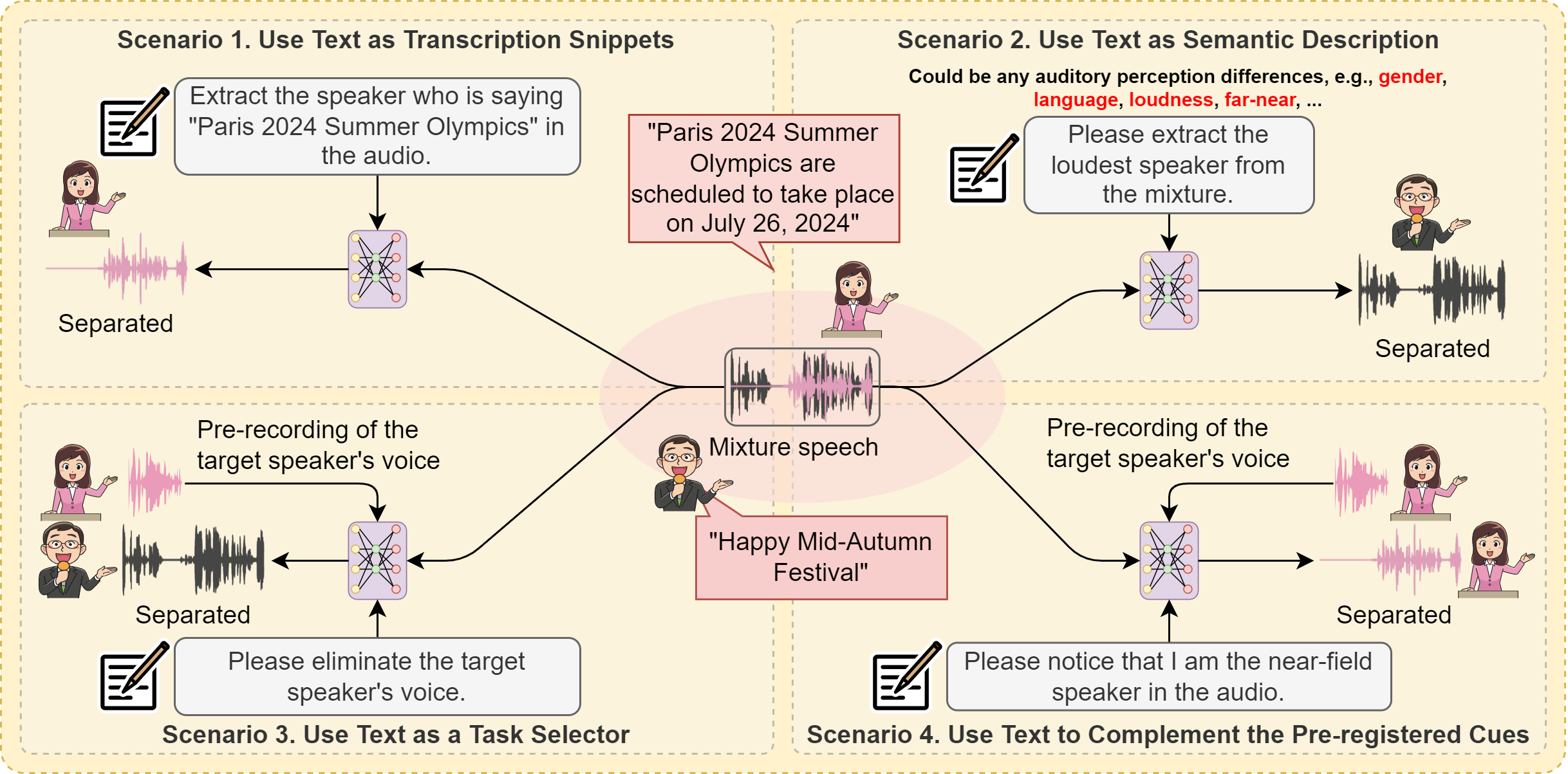}
    \caption{New application scenarios enabled by the proposed LLM-TSE model. The central part is a mixture audio sample where two speakers' voices overlap. The male speaker, although positioned at a greater distance from the microphone, has a voice with higher volume and is saying ``Happy Mid-Autumn Festival". In contrast, the female speaker is nearer to the microphone but speaks in a quieter tone, delivering the message ``Paris 2024 Summer Olympics are scheduled to take place on July 26, 2024". The illustration's four corners show the innovative application scenarios enabled by LLM-TSE.}
    \label{fig:scenario}
\end{figure*}

\section{Application Scenarios Enabled by LLM-TSE}
\label{sec:new_app}

The text-guided LLM-TSE model introduces a wide array of new application scenarios that significantly surpass the capabilities of existing target speaker extraction methods. As depicted in Figure~\ref{fig:scenario}, the center of the illustration presents an overlapping mixture speech from two speakers. The first is a man whose voice, despite being further from the microphone, is louder and is saying ``Happy Mid-Autumn Festival". The second is a woman whose voice is softer and is saying ``Paris 2024 Summer Olympics are scheduled to take place on July 26, 2024", although she is positioned closer to the microphone. In the four corners of Figure~\ref{fig:scenario}, we detail the novel application scenarios facilitated by this model, which are organized into four distinct types.

\subsection{Use Text as Transcription Snippets}
Humans utilize discernible cues in relatively clean speech segments to enhance the perception of highly corrupted speech segments~\cite{shinn-cunningham_selective_2008,popelka_hearing_2016}. Similarly, the LLM-TSE model can leverage distinguishable acoustic cues, in the form of transcription snippets, to facilitate speaker extraction. For instance, as illustrated in Figure~\ref{fig:scenario} Scenario 1, the LLM-TSE model allows us to extract a specific speaker from a mixed speech recording by using just a short transcription snippet, such as ``Extract the speaker who says `Paris 2024 Summer Olympics' in the audio." This command helps the model to identify and isolate the speech of the desired speaker.

\subsection{Use Text as Semantic Description} 
Apart from the above content-based cue, humans also employ many other perceptual cues based on the distinguishing characteristics between competing speakers, such as gender, language, loudness level, and reverberation in the audio signal. The LLM-TSE model enables users to incorporate such perceptual cues as text-based semantic descriptions to exert control over the process of target speaker extraction. Notably, these perceptual cues can be considered as independent pre-registered cues. For example, as depicted in Figure~\ref{fig:scenario} Scenario~2, we can instruct the model using natural language text such as ``Please extract the loudest speaker from the mixture," asking the model to identify and isolate the speech of the loudest person in the audio.

\subsection{Use Text as a Task Selector}
During a conversation involving multiple speakers, humans often switch their focus from one speaker to another. In addition, the speaker of interest at one moment may become a distraction at a later moment. In contrast to existing TSE systems that can only concentrate on a pre-registered speaker, the proposed LLM-TSE model empowers users with the flexibility to decide whether to retain or exclude the pre-registered speaker from the audio mixture, expanding beyond what is currently achievable with traditional TSE methods.
For instance, as shown in Figure~\ref{fig:scenario} Scenario 3, when provided with a pre-recorded speech to identify the speaker, we can command the model with ``Please eliminate the target speaker's voice" instead of extracting it. This instructs the model to suppress the identified speaker's voice, thereby allowing other speakers in the audio mixture to come to the forefront.

\subsection{Use Text to Complement Pre-registered Cues}
In conventional TSE systems, the voice of the target speaker is typically pre-recorded that may differ substantially from the actual deployment environments due to the change of acoustic environment or emotional state~\cite{zmolikova_neural_2023}. This discrepancy significantly affects the robustness of conventional TSE systems. In contrast, the proposed LLM-TTS model has the ability to compensate for these differences by providing complementary cues in addition to the pre-registered ones, such as the speaker's location, language, loudness level, etc. Consequently, it generates a more comprehensive and accurate representation of the target speaker that can significantly enhance the system's robustness. For example, as illustrated in Figure~\ref{fig:scenario} Scenario 4, after providing a pre-recorded voice to identify the speaker, we can enhance the model's accuracy by instructing it with a statement like, ``Please note that I am the near-field speaker in the audio." This extra information helps the model to refine its focus and extract the voice of the near-field speaker more effectively within the acoustic environment.

\begin{figure*}[ht]
    \centering
    \includegraphics[width=0.9\textwidth]{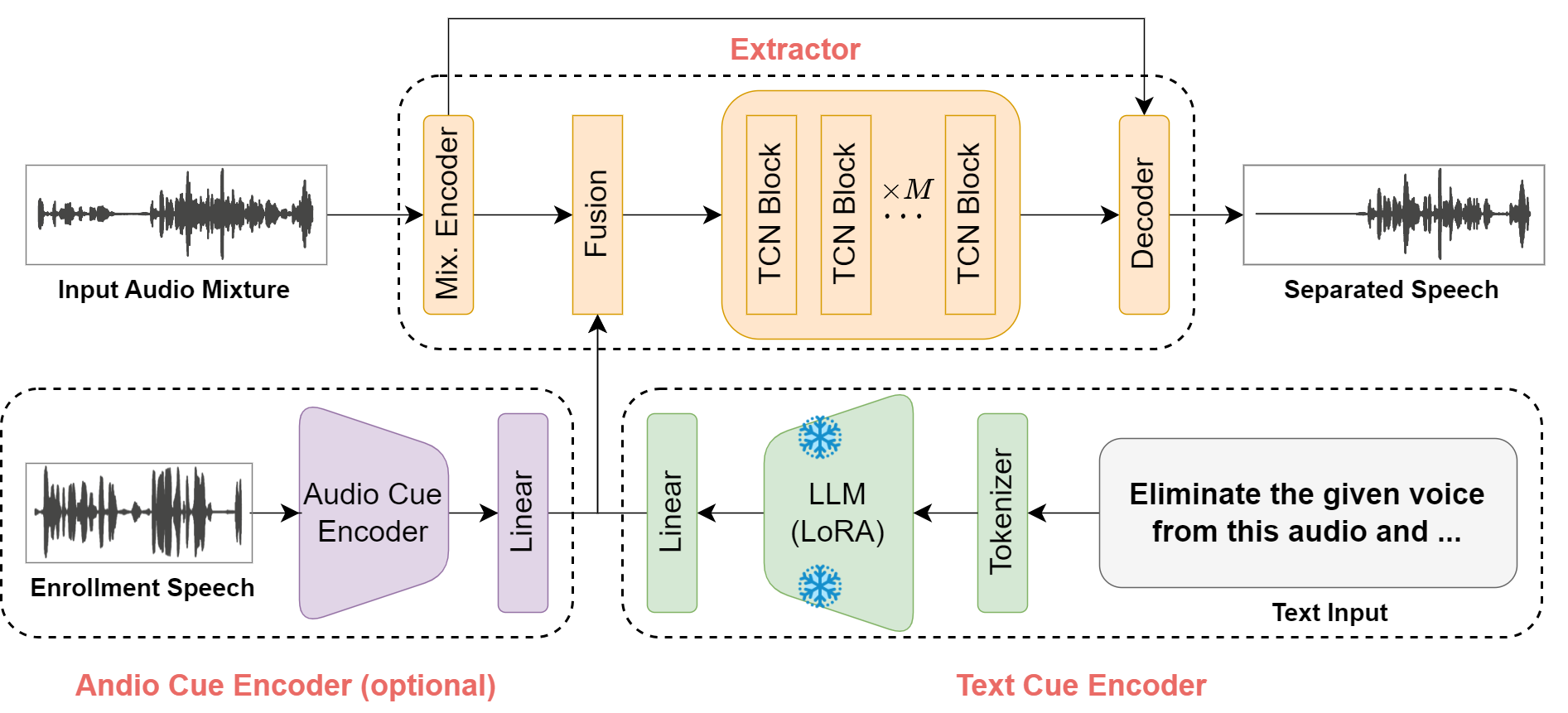}
    \caption{Overview of the proposed LLM-TSE model architecture. We use LoRA~\cite{hu_lora_2021} to fine-tune a small number of parameters of the LLM component.}
    \label{fig:model_arch}
\end{figure*}

\section{LLM-TSE Model}
\label{sec:model}

As shown in Figure~\ref{fig:model_arch}, our LLM-TSE model follows a processing pipeline of Encoding-Fusion-Extraction-Decoding. In the encoding phase, three distinct encoders are employed to convert the pre-recorded enrollment speech, nature language descriptions, and input mixture speech into corresponding embeddings, respectively. Then, leveraging the fused embeddings representing the pre-recorded enrollment speech and text cues, the extractor selectively extracts the desired speech source from the input mixture speech. Finally, the output feature representation obtained from the extractor is transformed into the time-domain and output as the extracted speech.

%Notably, the text and audio cue encoders, which can operate independently or collaboratively, are constructed to extract essential information from the input text and/or speech data. This information extraction enables the extractor module to identify and isolate the target speech from the mixture audio input. Contrastingly, the mixture encoder is only tasked with performing a time-frequency transformation. After the Encoding phase, the Fusion phase commences, merging the multiple embeddings into one cohesive form. The Extractor module then takes this fused embedding and the mixture as input, utilizing them to selectively extract the desired speech source from the combined speech features. Lastly, the Decoder stage restores these frequency-domain features to the time-domain as a extracted speech signal. This comprehensive process outlines our innovative approach to enhancing the precision and efficiency of target speaker extraction tasks.

\subsection{Mixture Encoder and Decoder}
The mixture encoder transforms the input audio mixture from the time domain to feature representation, which can be more effectively handled by the extractor~\cite{zmolikova_neural_2023}. This transformation is realized by convolving each audio frame of length $L$ with a set of $N$ 1-D convolution filters $\{ u_n(t) \}_{n = \{0 \dots N -1\}}$, which can be expressed as follows:
\begin{equation}
    \mathbf{X}(k, n) = \sum_{t=0}^{L - 1} x(t + kH) \cdot u_n(t), \quad n \in \{ 0, \dots, N - 1 \},
\end{equation}
where $x(t)$ is the input mixture signal, $k \in \{0, \dots, K - 1 \}$ is the frame index, $H$ is the hop size, and $\mathbf{X}(k, n)$ is the result of the convolution operation. Similarly, the decoder maps the extracted feature, denoted as $\mathbf{Y}(k, n)$, back to the time domain via a transposed 1-D convolution operation with $N$ synthesis filters  $\{ v_n(t) \}_{n = \{0 \dots N -1\}}$, and each has a length of $L$:
\begin{equation}
\hat{y} (t) = \sum_{k=0}^{K-1} \sum_{n=0}^{N-1} \mathbf{Y}(k, n) \cdot v_n(t - k H),
\end{equation}
where $\hat{y}(t)$ is the extracted audio signal in the time domain. 

\subsection{Text Cue Encoder}
We utilize the LLaMA-2 7B Chat LLM~\cite{touvron_llama_2023}, a dialogue-fine-tuned version of the LLaMA-2~\cite{touvron_llama_2023}, to obtain discriminative semantic embeddings from the user's text input. 
LLaMA-2 is pre-trained on a combination of natural language and programming language corpora in a self-supervised manner. LLaMA-2 7B Chat LLM is further fine-tuned from LLaMA-2 via instruction-tuning, which significantly enhances its performance on various reasoning and generation tasks. During our model training, instead of performing full fine-tuning on the adopted LLM text encoder, we adopt the parameter-efficient Low-Rank Adaptation (LoRA) technique~\cite{hu_lora_2021}. LoRA introduces a small set of parameters into the frozen LLaMA-2 7B Chat LLM, which are referred to as LoRA adapters. Specifically, one LoRA adapter is attached to each LLM layer, modifying its frozen parameter by adding a low-rank learnable matrix of the same size. In the proposed LLM-TSE model, we apply the LoRA adapters to only modify keys and queries in each self-attention layer. Ultimately, we only add 12\% more trainable parameters. This approach not only helps to prevent the overfitting problem that is often encountered with a small fine-tuning dataset but also improves the training efficiency.

\subsection{Audio Cue Encoder}
The primary role of the audio cue encoder is to encode the optional pre-registered speech into a discriminative speaker embedding. The first step in this encoder involves transforming the time domain input signal, using the above-mentioned learnable 1-D convolutional filters, into the feature representation. Following this transformation, we utilize a series of Temporal Convolutional Network (TCN) blocks~\cite{pandey_tcnn_2019,luo_conv-tasnet_2019} to extract speaker-related feature representation. These TCN blocks are designed to capture the temporal dependencies in the speech signal, which are crucial for distinguishing different speakers. Finally, we take the average along the temporal dimension to generate a speaker embedding vector, which effectively captures the unique vocal attributes of the pre-registered speech that can differentiate one speaker from others.

\subsection{Fusion Layer}
Here, we follow a simple concatenation approach to fuse the audio and text cues, which has been shown effective in many other TSE systems~\cite{zmolikova_speakerbeam_2019,ge_spex_2020,xu_spex_2020,transformerrealtime23}. Specifically, we transform the text cue and audio cue embeddings into the same dimensional through two linear projection layers, and then directly concatenate them to form a multi-modal representation. 

\subsection{Extractor}
The last part of our model is the target extractor, which serves to estimate the target signal. We adopt the widely used time-frequency masking-based extractor~\cite{luo_conv-tasnet_2019,isik_single-channel_2016}, and its operations can be summarized as follows:
\begin{equation}
\begin{split}
    \mathbf{M} & = \text{MaskNet}(\mathbf{Z}; \theta^\text{Mask}), \\
    \hat{\mathbf{Y}} & = \mathbf{M} \otimes \mathbf{X},
\end{split}
\end{equation}
where $\mathbf{Z}$ is the fused embedding generated from the fusion layer, $\text{MaskNet}(\cdot)$ is a TCN-based NN that estimates the time-frequency mask $\mathbf{M} \in {\mathbb{R}^{D \times N}}$ for the target speaker, where $D$ is the feature dimension of each time step. $\theta^\text{Mask}$ is the network parameter, and $\otimes$ denotes the element-wise Hadamard product. $\hat{\mathbf{Y}}$ is the estimated target speech signal in the frequency domain. 

\subsection{Loss function}
The parameters of the proposed LLM-TSE model are optimized by minimizing the following Scale-Invariant Signal-to-Distortion Ratio (SI-SDR)~\cite{roux_sdr_2019} loss function:
\begin{equation}
    \mathcal{L}^\text{SI-SDR} = - 10 \log_{10} \left(  \frac{\| \frac{\hat{\mathbf{y}} ^T \mathbf{y}}{\|\mathbf{y}\|^2} \mathbf{y} \|^2}{\| \frac{\hat{\mathbf{y}}^T \mathbf{y}}{\|\mathbf{y}\|^2} \mathbf{y} - \hat{\mathbf{y}}  \|^2}  \right).
\end{equation}
The SI-SDR loss is computed directly in the time domain, which forces the model to learn to precisely estimate the magnitude and the phase of the target speech signals.

\section{Experimental Setup}
\label{sec:exp_setup}
Our primary objective in this work is to integrate text-based cues to enhance the target speaker extraction systems. 
In this section, we initially delve into the method of simulating the overlapped mixture of speech data. Subsequently, we will explore the generation of text questions.

\subsection{Overlapped Speech Simulation}
Our experiment uses two speech datasets: LibriSpeech~\cite{panayotov_librispeech_2015} and Multilingual LibriSpeech (MLS) ~\cite{pratap_mls_2020}. LibriSpeech, a 1000-hour corpus of English audiobook speech, is known for its diverse speaker identities. MLS, an extension of LibriSpeech, adds multiple languages, including French, German, Spanish, etc. Due to it having too much data, we randomly select 400 speakers per language from MLS with up to 20 utterances each. We adhere to LibriSpeech's standard training, validation, and test set division. For MLS, we randomly assign 5\% of speakers from each language to validation and test sets, respectively, with the rest for training.

Our experiments cover a variety of attributes, including transcription snippets, gender, language, loudness, and far-near. For transcription snippets extraction, we only use the LibriSpeech dataset and the corresponding pre-extracted forced alignment~\cite{chodroff_montreal_2023} data~\footnote{\href{https://github.com/CorentinJ/librispeech-alignments}{https://github.com/CorentinJ/librispeech-alignments}} to identify the word timestamps from LibriSpeech. The remainder of the data for simulation is randomly selected from the LibriSpeech and MLS datasets. For generating the mixture speech, we adopt online simulation, generating the data needed for each iteration beforehand. The number of speakers in the mixture of speech is limited to two, stipulating that the two speakers have different attributes for gender, language, loudness, or far-near. When generating a mixture of speech for the loudness task, our signal-to-noise ratio is randomly selected from -3 dB to -2 dB and 2 dB to 3 dB. The other tasks span from -3 dB to 3 dB. In the case of the distance task, we include both near (target speaker) - far (interference speaker) and far (interference speaker) - near (target speaker) scenarios. For the other tasks, near and far combinations are randomized. Room dimensions are randomly selected from lengths of 9 to 11 m, widths of 9 to 11 m, and heights of 2.6 to 3.5 m.
The reverberation time ranges from 0.3 to 0.6 seconds. We use Pyroomacoustics~\footnote{\href{https://github.com/LCAV/pyroomacoustics}{https://github.com/LCAV/pyroomacoustics}} to generate Room Impulse Responses (RIRs), and the microphone's position is defaulted to the center of the room.
The sound source distance from the microphone varies between 0.3 to 0.5 m and 1.5 to 2.5 m for near or far fields, respectively. The angle ranges from 0 to 180 degree, and the sound source's height varies between 1.6 to 1.9 m.

The mixture and pre-registered speeches are set to a duration of 6 seconds, with a randomly determined overlap ratio between 40\% and 70\%. The pre-registered speech is randomly selected from the remaining target speaker's speech. If the training objective is to remove the target speaker, the other speaker's speech from the mixture serves as the training target.
We assume that each generated mixture speech sample should exhibit a distinguishable attribute throughout the training. All experimental data is sampled at 16,000 Hz to ensure high-quality audio.

\subsection{Text Generation}

We include three types of text to explore using LLMs to enrich target speaker extraction systems. We first create ten foundational question templates for each type of task. These templates will then be rephrased and expanded using ChatGPT-4-32K~\footnote{\href{https://platform.openai.com/docs/models}{https://platform.openai.com/docs/models}} to produce 100 diverse text prompts. The prompt of rephrase is: ``Keep it short, limit to 8 words. Feel free to vary sentence structures, but avoid duplications, and synonyms can be replaced. Imitate the tone of a casual conversation, don't be too rigid. Maintain the existing JSON format when outputting." We adopt a non-overlapped 80/10/10\% partitioning for training, validation, and testing sets. The text prompts used in the testing set are unseen during the training.

\subsubsection{Text as an Independent Extraction Cue}
In this type, the text is used as an independent extraction cue. The texts of this task are like: ``Extracting a voice with $\langle$specific characteristic$\rangle$ from a mixture of speech", e.g., scenarios 1\&2 in Figure~\ref{fig:scenario}. The text description outlines the features of the voice to be extracted, including the transcription snippets of the mixture of speech, the speaker's language, gender, loudness, and far-near. For the transcription snippet task, we use 100\% of the target speech text length as cues for training, testing with 50\%, 80\%, and 100\% of the target speech text length to evaluate generalizability. This setup is highly functional, i.e., by informing the system about the audible part of the speech, the system can utilize both semantic and acoustic information to track and extract the desired speaker. Note that the attributes utilized in this study are not exhaustive. In real-world situations, humans employ a variety of other cues, e.g., emotion or pitch, to extract the sound source of interest~\cite{haykin_cocktail_2005, shinn-cunningham_selective_2008}. However, exploring these additional cues extends beyond the scope of this current study and is reserved for future research.

\subsubsection{Text as a Task Selector}
We propose one task type where text can influence the system's output: target speaker extraction or removal. The text serves as a directive for the system to either extract a given speaker's voice or remove it from the mixture of audio. The generated texts are like ``please remove the given voice from this audio."

\subsubsection{Text as a Complement to Human Perception in the Voiceprint-Based Extraction System}
We integrate the human understanding and interpretation of the mixture of speech into the extraction process, which can significantly enhance the system's performance. Here, we cover all semantic types mentioned above, i.e., transcription snippets, gender, language, loudness, and far-near. The generated questions are like ``Extracting a speaker based on the given pre-registered speech, where the speaker possesses a $\langle$specific characteristic$\rangle$ within the mixture speech."

\subsection{Implementation Details}

\subsubsection{Model Architecture}
The LLM-TSE model incorporates a text cue encoder derived from the LLaMA-2 7B model, a transformer decoder architecture. We generate the text cue embedding using the averaging results of the outputs of the last four self-attention layers. Subsequently, a linear projection layer is employed to map its dimensions to match the embedding output of the audio cue encoder model. The construction of the audio cue encoder and extractor is built upon an open source code of the time-domain SpeakerBeam (TD-SpeakerBeam)~\footnote{\href{https://github.com/BUTSpeechFIT/speakerbeam}{https://github.com/BUTSpeechFIT/speakerbeam}}. The default model hyperparameters from TD-SpeakerBeam are employed in this process.

\subsubsection{Optimization}
We use the AdamW optimizer for optimization, with an initial learning rate of 1e-4, which has proven effective for various tasks in our preliminary experiments. Our model is trained using ten NVIDIA 3090 GPUs, each with a batch size of 1. For stable training, we employ gradient accumulation, with backpropagation performed every two interactions, culminating in a valid batch size of 40 per iteration. A linear warmup scheduler is used for the first 1000 iteration steps, during which the learning increases from 0 to 1e-4 and remains constant. This strategy aims to gradually prepare the model for more complex tasks and improve overall learning stability. Finally, based on our preliminary experiments on the current dataset, we use the gradient normalization with a value of 30. This operation controls the weight update step and prevents gradient explosion. 

\subsubsection{LoRA Adaptor}
We adopt the LoRA approach for efficient fine-tuning. The hyperparameters of the LoRA matrix, rank $r$, and scaling weight $\alpha$ are set to 16 and 16. The LoRA dropout is set to 0.05. These LoRA adaptors are applied to the linear projection layers of the query and key calculation in the self-attention layers.

\begin{table*}[t]
    \caption{Evaluation of SI-SDR (dB $\uparrow$) metric across different methods. For the transcription snippet task, we use 100\% of the target speech text as cues during training and test the model with a different amount of text transcriptions, including 50\%, 80\%, and 100\%.}
    \label{tab:results}
    \centering
    \small
    \resizebox{0.95\textwidth}{!}{%
    \begin{tblr}{c|cc|ccccccc}
    \toprule
    \SetCell[r=2]{c} Entry & \SetCell[c=2]{c} Type of Cue & & \SetCell[c=3]{c} Transcription Snippet & & & \SetCell[r=2]{c} \SetCell[r=2]{c} Gender & \SetCell[r=2]{c} Language & \SetCell[r=2]{c} Far-near & \SetCell[r=2]{c} Loudness \\ \midrule
        & Audio & Text & 50\% & 80\% & 100\% & & \\  \midrule
        Unproc. & \SetCell[c=2]{c} - & & \SetCell[c=3]{c} -0.02 & & & -0.02 & -0.03 & -0.01 & -0.10 \\ \midrule
        TD-SpeakerBeam & \Checkmark & \XSolidBrush & \SetCell[c=3]{c} 7.21 &  &  & 10.15 & 8.38 & 9.38 & 7.57 \\ \midrule
        % BSRNN & \Checkmark & \XSolidBrush & \SetCell[c=3]{c} 12.76 &  &  & 13.34 & 12.76 & 13.58 & 12.17 \\ \midrule
        \SetCell[r=5]{c} {LLM-TSE \\ (LoRA Adapters, \\ LLaMA-2 7B Chat)} & \Checkmark & \XSolidBrush & \SetCell[c=3]{c} 7.30 & & & 10.17 & 8.87 & 9.77 & 7.75 \\
        & \XSolidBrush & One-Hot & \SetCell[c=3]{c} No Support &  &  & 10.54 & 8.88 & 10.25 & 8.96 \\ 
        & \XSolidBrush & \Checkmark & 2.70 & 3.97 & 7.48 & 10.40 & 9.38 & 10.57 & 8.89 \\ 
        & \Checkmark & One-Hot & \SetCell[c=3]{c} No Support & &  & 10.62 & 10.18 & 10.32 & 8.99 \\ 
        \SetRow{gray9} & \Checkmark & \Checkmark & 7.96 & 9.81 & 10.05 & 10.87  & 9.72 & 10.66 & 9.41 \\ \midrule
        % [10.622366791598376, 10.186836328931411, 10.322450196761626, 8.98620811782299]
        % \SetCell[c=10]{c}{\small \textit{\color{gray} Ablation Studies}} \\ \midrule
        % \SetCell[r=1]{c} \textit{all} \\
        \SetCell[r=2]{c} {No LoRA Adapters \\ (only Linear Projection) } & \XSolidBrush & \Checkmark & 1.66 & 3.38 & 5.38 & 8.76 & 7.38 & 8.45 & 5.46 \\
        & \Checkmark & \Checkmark & 4.85 & 7.60 & 7.98 & 9.02 & 7.97 & 8.67 & 7.11 \\ \midrule
        \SetCell[r=2]{c} {Use Vicuna-7b-v1.3 \\ (\cite{zheng_judging_2023})} & \XSolidBrush & \Checkmark & 2.23 & 3.31 & 8.79 & 9.44 & 8.29 & 9.27 & 5.75  \\
         & \Checkmark & \Checkmark & 7.41 & 9.05 & 9.35 & 10.15 & 9.01 & 9.94 & 6.47 \\
    \bottomrule
    \end{tblr}
    }
\end{table*}

\begin{figure*}[t]
    \centering
    \includegraphics[width=\linewidth]{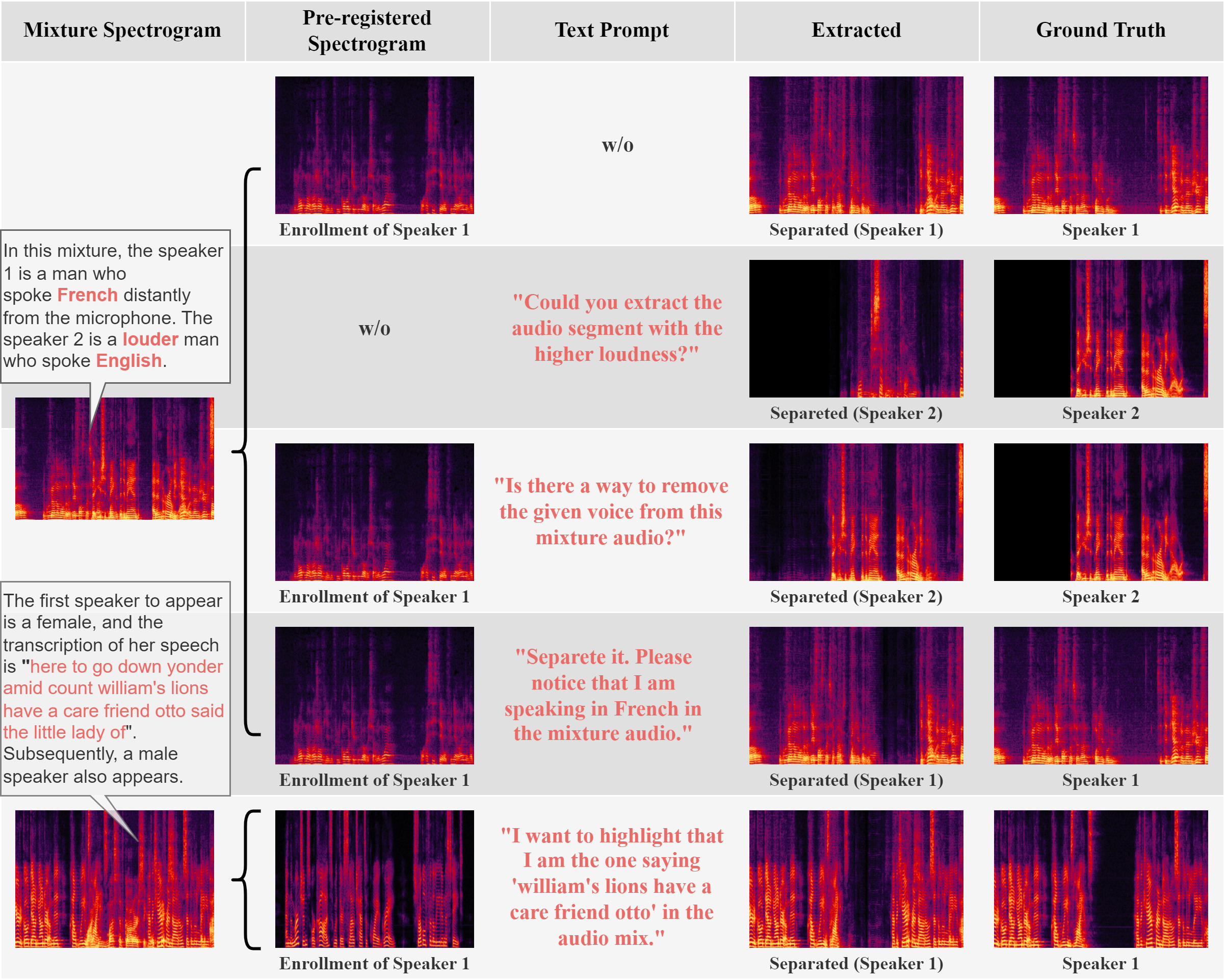}
    \caption{Samples generated from the proposed LLM-TSE model. The text box contains information about the input audio mixture. The term ``w/o" indicates the absence of a certain input.}
    \label{fig:samples}
\end{figure*}

\section{Experimental Results}
\label{sec:exp_result}
In this section, we evaluate the performance of the LLM-TSE model on the mixture overlapped speech dataset. Section~\ref{sec:independent} showcases how the LLM-TSE model significantly advances target speaker extraction by utilizing text as an independent cue. Section~\ref{sec:task_selector} details the model's use of text to selectively control the speech separation process. Performance enhancements from text complementing pre-registered cues are examined in Section~\ref{sec:complement}. Finally, Section~\ref{sec:complement} discusses the impact of employing different text encoders on the system's efficacy.

\subsection{Efficacy of Using Input Text as Independent Cues}
\label{sec:independent}
Table~\ref{tab:results} demonstrates a notable performance enhancement when text alone is employed as an extraction cue, compared to unprocessed mixture speech. The proposed LLM-TSE model is built on TD-SpeakerBeam~\cite{delcroix_improving_2020}, a state-of-the-art (SOTA) open-source target speaker extraction model. Compared to TD-SpeakerBeam, the only modification in the LLM-TSE model is the additional text encoder. This enhancement is further corroborated by Figure~\ref{fig:samples}.
These findings suggest that the LLM-TSE model effectively interprets the provided text descriptions, which fundamentally serve as human interpretations of auditory object differences within a speech mixture. This innovative strategy represents a significant leap in harnessing natural language processing techniques for complex auditory tasks, thereby enhancing the scope of potential applications for speaker extraction methodologies.

\subsection{Compared with One-Hot System}
We notice that some attribute-based questions (such as language, gender, loudness, and distance) can be encapsulated into a one-hot representation, which can be used as a baseline to assess the comprehension capabilities of LLMs. We can notice that LLM-based system has achieved performance that is very close to that of the one-hot system, which shows that for any questions of these attribute classes, the LLM component can successfully understand natural language descriptions. However, we must acknowledge the limitations inherent in using one-hot representations:
\textbf{1)} One-hot representations are only capable of expressing attributes with distinct classifications, for instance, language, gender, and loudness. If we want to employ other cues, like transcription snippets, one-hot representations prove insufficient.
\textbf{2)} One-hot representations lack adaptability. LLMs can aid the target speaker extraction system in interpreting user text inputs, thus facilitating the injection of more generic and diverse semantic cues. For example, the input of LLM-TSE can be effortlessly extended to support open-ended questions, such as ``isolate the speaker based on the 3-4 second segment in the mixed speech," a task beyond the capacity of one-hot representations.

\subsection{Efficacy of Using Input Text as Task Selector}
\label{sec:task_selector}
In this experiment, we inspect whether our model can control the training targets of the separation system using natural language. The corresponding textual queries could resemble ``Is there a way to remove the given voice from this mixture audio?" In Figure~\ref{fig:samples}, we illustrate the capacity of our system to determine whether to extract or suppress the sound source corresponding to the provided pre-registered speech when using text descriptions. Notably, the samples displayed in the third row exemplify this capability, as they successfully suppress the target sound source associated with the pre-registered speech. Our explorations in this area are somewhat limited at this stage. 
More broadly, we expect these controls to be configured with greater flexibility in future, e.g. manipulating the degree of reverberation in the extracted speech (since individual preferences for reverberation vary), or dictating the impact range of the separation system (to avoid unnecessary non-linear-processing distortion). We intend to delve deeper into these aspects in our future work.

\subsection{Efficacy of Using Input Text to Complement the Pre-registered Cues}
\label{sec:complement}
Pre-registered speech primarily only encodes the speaker's vocal characteristics regardless of any time or acoustic environmental context. We aim to introduce this contextual information into the target speaker extraction system utilizing text descriptions. For this purpose, a typical text description is like: ``Separate the target speaker's audio based on the provided pre-registered speech as a reference, bearing in mind that I am the speaker who employs a louder tone in the mixed speech". The relevant experimental outcomes are presented in the middle section of Table~\ref{tab:results}.
Upon integrating descriptions delineating auditory object differences, we observe a significant improvement in system performance. This enhancement is particularly prominent in the ``loudness" task, where the dataset contains a pronounced loudness disparity between the two sound sources. The challenge posed by identifying the target speaker using only the pre-registered speech is substantially mitigated upon implementing our approach, producing the most substantial performance increase within this task.

\subsection{Ablation Studies on Text Encoder Selection}
Here, we present the results of a sequence of ablation experiments executed on the text encoder component. The outcomes are summarized at the bottom of Table~\ref{tab:results}. At the outset, we assess the functionality of the text cue encoder in the absence of the LoRA adaptors, where only the projection layer of the LLM model is permitted to train, effectively freezing all other parameters of the LLM. This configuration aims to determine if the LLM's generic understanding of diverse text corpora could offer sufficient discriminative information. However, our findings suggest that relying solely on embeddings, derived from the LLM's interpretation of various text descriptions, is insufficient to accomplish the task whether an audio encoder is integrated into the system or not. In subsequent experiments, we employ the Vicuna 7B model~\cite{zheng_judging_2023} as our text encoder. This model, which is fine-tuned on data from ``shareGPT.com" and based on the LLaMA-v1 model, exhibits marginally inferior performance in natural language benchmark tasks compared to the LLAMA-2 7B Chat. Further, the Vicuna model underperforms in our target speaker separation task compared to the LLAMA-2 7B Chat. This observation supports the premise that employing a more powerful LLM as a text cue encoder can significantly enhance the discriminative capabilities of the overall system.

\section{Conclusion and Future Works}
\label{sec:con}

In this work, we explore a novel paradigm for target speaker extraction, namely LLM-TSE, a significant departure from previous methodologies. The LLM-TSE approach uniquely introduces nature language descriptions to provide useful speaker extraction cues, effectively enhancing the feasibility, controllability, and performance of current TSE models.  
As indicated by our experimental results: \textbf{1)} Text proves its capability to act as a standalone extraction cue, potentially addressing the privacy issues inherent in predominant voiceprint-based target speaker extraction systems, whilst being very cheap to obtain. \textbf{2)} The use of text input allows the model to either extract or eliminate a target speaker, overcoming the constraints associated with extracting only pre-registered voices. \textbf{3)} Finally, by informing TSE models about the speaker's current state, text can help tackle intra-speaker variability, thereby enhancing the effectiveness of speaker extraction. In summary, our proposed paradigm signifies an important advancement for target speaker extraction systems, extending accessibility and improving performance. Not only does it provide a fresh perspective on the extraction process, but it also lays the groundwork for potential future studies on the cocktail party problem.

While these initial results are encouraging, many challenges remain. In the future, we aim to incorporate a range of mutually exclusive or non-exclusive auditory attributes (e.g., pitch, timbre, and speech speed rate), open-ended text descriptions, and develop the capability for multi-round target speaker extraction.

\bibliographystyle{IEEEtran}
\bibliography{main}

\end{document}